\documentstyle[12pt]{article}
\begin{document}
\begin{center} {\Large \bf
Initial Data Set For Cosmology: Application to Matching Condition}

Houri Ziaeepour\\ESO, Schwarzchildstrasse 2, 85748, Garching b. M\"{u}nchen, Germany\\
Email: {\tt houri@eso.org}
\end{center}

\begin {abstract}
In Einstein theory of gravity the initial configuration of metric field and its 
time derivative are related to matter configuration by four equations called 
constraints. We use the method of conformal metrics (York Method) to solve 
constraints and find an analytic set of consistent initial data for linearized 
Einstein field equations in a  perturbed constant curvature space-time.
They are explicitly covariant and more compact than decomposition of quantities 
to scalar, vector and tensor. This method is independent of type and physics of 
matter fields and is extendable to higher-order perturbative calculations. 
As an application example, we apply this method to two commonly used matching 
conditions during a phase transition and compare and interpret the results. 

\begin {description}
\item {Keywords:}\\
\hfill cosmology -- relativity:initial conditions -- cosmology:\\
\hfill perturbations -- relativity: conformal metric
\end {description}

\end {abstract}

\section{Introduction}
By a set of initial conditions, we mean a configuration for matter, radiation,
and metric fields, and their time derivatives on a given initial 3-space. This 
information is necessary for solving Einstein equations.  
The task of defining this configuration is not trivial \cite {Linch 44}, 
\cite {ADM 62}. In contrast to Newtonian theory, due to diffeomorphism 
gauge symmetry, the 
initial value of the metric and its time derivative can not be specified in an 
arbitrary way. The configuration of matter fields partially determines them.\\ 
To define initial conditions in general relativity properly, one must define 
a time slicing of the space-time i.e. a diffeomorphism between the 
space-time and a 3+1 manifold \cite {yorkconf}, \cite {york1}, 
\cite{gravmath}, \cite {texture}. From this operation, in addition to 
evolution equations for geometrical quantities, one obtains the restriction 
of the Einstein equations on the space-like 3-space component of the 3+1 
manifold. These equations don't evolve with time and for this reason 
they are called constraints \cite {ADM 62}. The initial condition for 
geometric and matter fields is defined as their configuration on a space-like  
3-space. As constraints are not dynamical equations, it is necessary and 
sufficient that field configuration on the initial 3-space satisfy them.\\
In cosmology, specially for studying the evolution of small perturbations, it 
is customary to decompose Einstein equations, as well as energy-momentum 
tensor, to scalar, vector, and tensor components. Evolution equations and 
constraints for each type of fluctuations are solved separately. This procedure 
drops some of components from constraints and makes them easier to solve, 
specially when only one of components, usually scalar component, is  
studied. Nevertheless, evolution equations and constraints remain coupled and 
one has to solve them together. Moreover, the resulting 
equations are not usually explicitly covariant (although it is possible to 
perform the decomposition in a covariant manner \cite{covardecomp}).\\
Having an exact solution of constraints is also important in numerical 
solution of evolution equations. Usually, it is very difficult to keep the 
conservation of energy-momentum in a numerical calculation. The exact solution 
of constraints assures that the initial data satisfies the conservation laws. 
It also can be used at each step of calculation for checking/correction of 
conservation violation.\\
The mathematical aspects of the initial conditions for the Einstein field 
equations have been studied and clarified in an outstanding work by J.W. 
York on the initial condition problem and its relation to conformal gravity 
\cite {yorkconf}, \cite {york1}. This formalism separates initially constrained
components of metric and extrinsic curvature, and allows a more 
detailed insight to the physical nature of unconstrained components without 
knowing anything about the matter content of the theory.\\ 
Here we apply this method to linearized Einstein equations, solve constraints 
analytically, and find a 
consistent set of initial conditions for perturbed constant curvature 
cosmogonies. The results are independent of details of the physical model 
under consideration. As an application example, we use the results of this 
method to discuss matching during a phase transition.\\ 
We first briefly remind the mathematical formulation of the
problem. Then, we explain analytical solution of constraints for flat and 
constant curvature perturbed space-times and finally we apply the solution 
of constraints to matching conditions on a phase transition surface and we 
obtain the initial values allowed for unconstrained components.

\section{3+1 Gravity}
We assume that by a diffeomorphism transformation, the space-time is divisible 
to a space-like 3-manifold and a one-dimensional time like space (curve). 
The general form of the metric is: \footnote {Greek indexes correspond to 
space-time and Latin indexes to 3-space. We use units where $c = 1$ 
and $\hbar = 1$. The notations ",", "$|$", and ";" are used respectively for 
partial derivative, covariant derivative in 3-space and covariant derivative 
in space-time.}\\ 
\begin{equation}
ds^2 = -{\alpha}^2 d{\eta}^2 + g_{ij}(dx^i + {\beta}^i d\eta)(dx^j + 
{\beta}^j d\eta) \quad \quad dt^2 \equiv {\alpha}^2 d{\eta}^2 \label {metric}
\end{equation}
$\eta$ is the conformal time. ${\alpha} (\eta, x)$ and $\beta (\eta, x)$ are 
called lapse function and shift vector 
(in 3-space). The 3-tensor $g_{ij}$ is the induced metric of the 3-space.\\  
Einstein field equations can be expressed 
by a set of first-order differential equations which depend on the field set 
$\{{\alpha}, {\beta}^i, g_{ij}, K_{ij} \}$. The extrinsic curvature $K_{ij}$ is 
defined as the covariant derivative of the normal vector to the 3-space 
$K_{ij} = - n_{i;j} = -\alpha ^{(4)}\Gamma^0_{ij}$ \cite {grav}. The vector 
$n^\mu$ is the normal unit vector of the 3-space.
With respect to these fields, Einstein field equations take the following 
form \cite{numgrav}:
\begin{eqnarray}
\frac {\partial g_{ij}}{\partial \eta} & = & -2{\alpha}g_{ik}{K_j}^k + 
{\beta}^kg_{ij,k} + g_{ik}{\beta}^k_{,j} + g_{jk}{\beta}^k_{,i}. \label{gevol}
\\ 
\frac {\partial {K_i}^j}{\partial \eta} & = & {\beta}^k {K_i^j}_{,k} + 
{K_k}^j{{\beta}^k}_{,i} - {K_i}^k{{\beta}^j}_{,k} - {{\alpha}^j}_{|i} + 
\nonumber \\ 
 &   & {\alpha} (K{K_i}^j + {^{(3)}R_i}^j + 8{\pi} G g^{jk}T_{ik}). \label {kevol}
\end{eqnarray}
\begin{eqnarray}
^{(3)}R - {K_i}^j{K_j}^i + K^2 & = & 16{\pi} G T_*^* . \label {enrcons}\\ 
{K_i}^j_{|j} - K_{|i} & = & - 8{\pi} G T_i^*. \label {momcons}
\end{eqnarray}
The quantities $T_i^* = T_{\mu\nu}n^{\mu}B_i^{\nu}$ and $T_*^* = 
T_{\mu\nu}n^{\mu}n^{\nu}$ can be interpreted respectively as projection of 
energy-momentum flux on the 3-space and on its normal. 
$B^i_{\mu}$ is the projection operator on the 3-space, $B^i_{\mu}n^\mu = 0$. 
For metric (\ref {metric}):
\begin {eqnarray}
n^\mu & = & {\alpha}^{-1}(1, {\beta}^i). \\
T_i^* & = & {\alpha}^{-1} (T_{0i} - T_{ik}{\beta}^k). \\
T_*^* & = & {\alpha}^{-2}(T_{00} - 2 T_{0i}{\beta}^i + T_{ij}{\beta}^i
{\beta}^j). \label {tproj}
\end {eqnarray}
Equations (\ref{gevol}) and (\ref{kevol}) are dynamical equations for the evolution of the 
3-space. Equations (\ref {enrcons}) and (\ref {momcons}) don't have explicit 
time dependence. They are constraints. 
If they are satisfied by the metric and extrinsic curvature at the 
initial time $t_0$ on the 3-space, they stay valid for ever.\\ 
Gauge symmetry allows to fix arbitrarily the value of ${\alpha} (t, x)$ 
and $\beta (t, x)$. In synchronous gauge that we will use in this letter, 
${\alpha}$ depends only on $t$ and $\beta^i (t , x) = 0$. The gauge symmetry 
assures that these relations remain valid after the evolution of the dynamical 
system. In fact, 4 of 16 equations (\ref {gevol}) - (\ref {momcons}) are 
automatically satisfied due to Biancci identities. This results to an equal 
number of equations and fields. The physical reason behind existence 
of constraint equations is the gauge symmetry. A system with gravitational 
interaction is a constrained dynamical system and the initial value of $g_{ij}$ 
and $K_{ij}$ can not be arbitrarily chosen. The initial 
$g_{ij}$ and $K_{ij}$ fields configuration on the initial 3-space depends on 
the matter configuration and must satisfy constraints (\ref {enrcons}) and 
(\ref {momcons}).

\section{Conformal Metric Method}
In \cite{york2} and \cite {york1} it has been proved that unconstrained 
degrees of 
freedom of a gravitational system on a space-like 3-space are conformally 
equivalent. This allows to separate constrained and unconstrained components 
of $g_{ij}$ and $K_{ij}$. The extrinsic curvature tensor of the 3-space can 
be decomposed to traceless-transverse, longitudinal (traceless), and trace 
components: 
\begin {eqnarray}
K_{ij} & = & S_{ij} + (LW)_{ij} + \frac {1}{3} K g_{ij}. \label {kdecom}\\
(LW)_{ij} & = & W_{i|j} + W_{j|i} - \frac {2}{3} g_{ij} W_{|c}^c. 
\label {lw}
\end {eqnarray}
The scalar $K$ is the trace of $K_{ij}$ and $S_{ij}$ is its traceless-
transverse component ($S_{ij}^{|i} = 0$). $W^i$ is a 3-vector field that 
generates the longitudinal (traceless) part of $K_{ij}$.\\ 
The conformal transformation of a 3-metric $g_{ij}$ is defined as 
$\bar {g}_{ij} = {\phi}^4 g_{ij}$. We call $g_{ij}$ the base metric and $\bar {g}_{ij}$ the physical metric which satisfies Einstein equations. Under 
this transformation, (\ref {kdecom}) and (\ref {lw}) keep their form. 
The only difference between base 
or physical extrinsic curvature decomposition is the use of base or physical 
metric in the contraction of indexes. Therefore: 
\begin {eqnarray}
\bar K_{ij} & = & \bar S_{ij} + (\bar LW)_{ij} + \frac {1}{3} K \bar g_{ij}. 
\label {kbdecom}\\
(\bar LW)_{ij} & = & W_{i|j} + W_{j|i} - \frac {2}{3} \bar g_{ij} W_{|c}^c. 
\label {lbw} \\
\bar \Gamma^i_{jk} & = & \Gamma^i_{jk} + 2 \phi^{-1}(\delta^i_j \phi_{|k} + 
\delta^i_k \phi_{|j} - g_{jk}\phi^{|j}).
\end {eqnarray}
The trace $K$ and the vector potential $W^i$ are considered as invariant under 
this transformation. The traceless-transverse component $S_{ij}$ is chosen to 
transform as:
\begin {equation}
\bar {S}_{ij} \equiv {\phi}^{-2} S_{ij}. \label {sijtrans}
\end {equation}
(When only one initial 3-space is considered, the conformal transformation 
of $S_{ij}$ is arbitrary. 
However, it has been recently shown that by assuming two infinitesimally 
close 3-spaces and a mapping between them, this transformation rule is 
imposed by Einstein equations \cite {initswandich}).
Even though the choice for transformation of $T^{*i}$ and $T_*^*$ is not 
unique, following choices guarantee the existence of a non-spacelike energy 
flow from the 3-space \cite {york1}: 
\begin {eqnarray}
\bar T^{*i} & = & {\phi}^{-10} T^{*i}. \label {titrans}\\
\bar T_*^* & = & {\phi}^{-8} T_*^*. \label {ttrans}
\end {eqnarray}
With these relations, the longitudinal component must transform as:
\begin {equation}
(\bar LW)^{ij} = {\phi}^{-4}(LW)^{ij}. \label {lwtrans}
\end {equation}
$\bar {g}_{ij}$ and $\bar {K}_{ij}$ satisfy the constraint equations 
(\ref {enrcons}) and (\ref {momcons}) if $W^i$ and $\phi$ satisfy following 
equations:
\begin {eqnarray}
[{\phi}^6 (LW)^{ij}]_{|j} & = & \frac {1}{3} {\phi}^6 K_{|i} - 8 \pi T^{*i}.
\label{weq}\\
-8{\triangle} \phi & = & - R\phi + M_{TT} \phi^{-7} + 2 M_{TL} \phi^{-1} +
 \nonumber \\ 
 & & (M_{L} - \frac {2}{3} K) \phi^{5} + 16 \pi G T_*^* \phi^{-3}. 
\label{phieq}\\
M_{TT} & = & g_{ac} g_{bd} S^{ab} S^{cd}. \label {mtt}\\
M_{TL} & = & g_{ac} g_{bd} S^{ab} (LW)^{cd}. \label {mtl}\\
M_{L} & = & g_{ac} g_{bd} (LW)^{ab} (LW)^{cd}.\label {ml}
\end {eqnarray}
All indexes and derivatives are contracted with the base metric $g_{ij}$. 
Equations (\ref {weq}) and (\ref {phieq}) are obtained from constraint 
equations (\ref {enrcons}) and (\ref {momcons}). These equations show that 
$\phi$ and $W^i$ are the real constrained degrees of freedom of the initial 
configuration of fields. The unconstrained degrees of freedom are $g_{ij}$, 
$S_{ij}$ and $K$ and they are all conformally related to their physical 
counterpart. Note also that this formulation of the initial value problem is 
gauge independent.\\
The real difficulty in solving equations (\ref {weq}) and (\ref {phieq}) 
is that they are highly nonlinear and in general completely coupled. This 
fact has encouraged the use of other methods, specially for nonlinear numerical 
calculations \cite{numgrav}.

\section {Initial Data for Linearized Einstein Equations in Flat Cosmogony}
We consider an initial equal-time 3-space $\Sigma$ in a flat universe with 
small perturbations (it is always possible to redefine coordinates such that 
the 3-space become equal-time). A priori $\Sigma$ can be any hypersurface, 
but interesting 
cases are those with constant $K$ or $K = K_{homo} + \delta K$, such that 
$K_{|j} = (\delta K)_{|j}$ and $K_{homo}$ is regarded as the trace of 
$K_{ij}$ for a 3-space 
obtained from mapping $\Sigma$ to the homogeneous background manifold. In fact, 
$K$ is an appropriate quantity to time label a spacelike 3-space because it 
has a non-negative derivative in the direction of timelike orthogonal vector 
to the 3-space \cite{yorkconf}. Other physically interesting 3-spaces e.g. 
one with constant energy in a perturbative theory are close to a constant $K$ 
3-space. Here we only consider these types of 3-spaces for imposing initial 
conditions.\\
We take the base metric to be the metric in a flat homogeneous 
Freedman-Lema\^itre cosmology, i.e.:
\begin {equation}
g_{ij} = {\alpha}^2 (t) {\delta}_{ij}. \label {baremet}
\end {equation}
This choice is physically motivated, because for a homogeneous universe, this 
metric satisfies Einstein equations. In addition, it simplifies all calculations.\\
With this choice, in a flat universe with small perturbations, the conformal 
factor will be close to 1:
\begin {equation}
\phi ({\bf x}, t) = 1 + \delta \phi ({\bf x}, t) \label {delphi}
\end {equation}
For linearized Einstein equations in synchronous gauge, the 
physical metric $\bar g_{ij}$ is:
\begin {eqnarray}
\bar g_{ij} & = & {\alpha}^2 (t) ({\delta}_{ij} + h_{ij}) \nonumber \\ 
& = & {\alpha}^2 (t) {\phi}^4 ({\bf x}, t) {\delta}_{ij} = {\alpha}^2 (t) 
{\delta}_{ij} (1 + 4 \delta \phi ({\bf x}, t)). \label {gdef}
\end {eqnarray}
\begin {equation}
h_{ij} = 4 {\delta}_{ij} \delta \phi. \label {hij}
\end {equation}
This special form of metric fluctuation is not a restriction of 
perturbations to scalars and is not equivalent to Newtonian gauge because its 
application is uniquely on the initial 3-space. As we have seen in the 
previous section, the transverse-traceless component of the extrinsic curvature 
is an unconstrained quantity. If on the initial 3-space it is not zero, this 
pure tensorial curvature perturbation will contribute to $g_{ij}$ evolution 
(see (\ref {gevol})) and induces a purely tensorial metric perturbation.\\
Another important point is that it is always possible to choose a coordinate 
system on the 3-space with a metric like (\ref {gdef}). It is well known that 
synchronous gauge does not define the gauge completely and 
allows a redefinition of the coordinates which preserves the gauge. In a  
gauge preserving coordinate transformation like the following:
\begin {equation}
t' = t \quad \quad \mbox {and} \quad \quad x'^i = x^i + {\partial}^i \psi + 
{\varepsilon}^{ij} {\omega}_j
\end {equation}
the arbitrary fields $\psi$ (local translation) and ${\omega}_i$ (local 
rotation) can be chosen such that $\bar g_{ij}$ gets the form given in 
(\ref {gdef}). After calculating the initial condition, one can return to 
the previous coordinates or to any other gauge.\\ 
From definition of projected energy-momentum tensor Equation.(\ref {tproj}), in 
synchronous gauge $\bar T^{*i} = \bar T_0^i$. In an approximately homogeneous 
and flat universe with small perturbations, $\bar T_0^i = {\mathcal {O}} (1)$ 
in the comoving frame. Therefore, $\bar T^{*i} = 0 + \delta \bar T^{*i}$. 
Regarding (\ref {titrans}), at first-order of approximation, 
$T^{*i} = \bar T^{*i}$. From (\ref {momcons}), one can conclude that at  
zero-order $\bar K_{i|j}^j = 0$. This means that at zero-order, the physical 
extrinsic curvature is transverse, and therefore, in spaces with small 
perturbations, the value of its longitudinal part must be small. 
Equation (\ref {lwtrans}) shows that the longitudinal part 
of the base extrinsic curvature also must be small.\\
For small perturbations, equation (\ref {weq}) becomes:
\begin {equation}
[(1 + 6 \delta \phi) (LW)^{ij}]_{|j} = - 8 \pi G T^{*i} + \frac {1}{3} 
(1 + 6 \delta \phi) K^{|i}.\label {weq1}
\end {equation}
The term $\delta \phi (LW)^{ij}$ is of second order and one expects that for 
smooth fluctuations, its derivative must be negligible with respect to 
$(LW)^{ij}_{|j}$. As mentioned earlier, we assume that $K^{|i} = 
{\mathcal {O}} (1)$. Therefore, if: 
\begin {equation}
(\delta \phi (LW)^{ij})_{|j} \ll (LW)^{ij}_{|j}, \nonumber
\end {equation}
(\ref {weq1}) reduces to:
\begin {equation}
(LW)^{ij}_{|j} = - 8 \pi G T^{*i} + \frac {K^{|i}}{3} \approx - 8 \pi G 
\bar T^{*i} + \frac {K^{|i}}{3}.\label {weq2}
\end {equation}
The solution of this equation in the case of a flat base metric is trivial. 
With a variable change $x'^i = \alpha (t) x^i$, (\ref {weq2}) becomes:
\begin {equation}
{\partial}'^j{\partial}'_j W^i + \frac {1}{3}{\partial}'^i{\partial}'_j W^j = - 8 \pi G T^{*i}({\bf x'}, t) + \frac {{\partial}'^i K}{3}. \label {weq3} 
\end {equation}
$\partial'$ means partial derivative with respect to $x'$. The solution in real 
space is:
\begin {eqnarray}
W^i({\bf x},t) & = & \frac {1}{(2 \pi)^3}\int d^3 {\bf k} e^{-i\alpha (t)
{\delta}_{ij}k^ix^j}\frac {2 \pi G}{k^4} \nonumber \\
 & & [k^2 (4 T^{*i}({\bf k}, t) - \frac {k^i \delta K}{8 \pi G}) - 
k^i k_j T^{*j}({\bf k}, t)]. \label {wsol}
\end {eqnarray}
For solving equation (\ref {phieq}), one needs also $S_{ij}$ the transverse 
part of the base extrinsic curvature. It is an unconstrained component and 
must be chosen according to the physics of the  system. 
To understand better the physical r\^ole of $S_{ij}$, note that its 
definition:
\begin {equation}
S_{ij}^{|i} = 0 \label {sttdef}
\end {equation}
can have non-trivial solutions. It has been proven that this equation allows a 
solution with a singularity called monopole solution (a black hole, a point 
like object of finite mass, or a geometrical singularity at origin) 
\cite{yorkhole}:
\begin {eqnarray}
S^{ij} & = & \frac {3}{2r^2} (P^i u^j + P^j u^i - (g^ {ij} - u^i u^j) P^c u_c) 
+ \nonumber \\
& & \frac {3}{r^3} ({\epsilon}^{imn}S_m u_n u^j + {\epsilon}^{jmn}S_m u_n u^i). 
\label {sttsol}
\end {eqnarray}
The vector $\vec r$ is the radial vector through origin and $u^i$ is the 
radial unit vector, $P^i$ and $S^i$ are respectively the linear and angular 
3-velocity of the singular point. This solution can be extended to a solution 
with $N$ singularity at $\vec r^{(i)}, i = 1\cdots N$ \cite{yorkhole} 
\cite {nhole}. In fact, this is a solution for (\ref {weq2}) in 
vacuum. However, in this case, $(LW)^{ij}$ is transverse 
and it is possible to add (\ref {sttsol}) to any solution of (\ref 
{sttdef}) or in general to $K_{ij}$. The existence of this solution 
reflects the effect of a velocity field on the extrinsic curvature and on 
the evolution of the metric (equation (\ref {gevol})). Note also that 
$T^{\mu\nu}$ does not fix the velocity field. This make a solution of 
type (\ref {sttsol}) independent of (\ref {wsol}). In decomposition method, the 
linear and angular velocity fields appear in the decomposition of 
$T_{\mu \nu}$ and their geometrical r\^ole is not as evident as here.\\
The solution (\ref {sttsol}) can be generalized to a non-singular matter 
distribution:
\begin {eqnarray}
S_{vel}^{ij} & = & {\int}_{V - \{x\}} \sqrt {g} d^3 {\bf x'} \frac {3}{2|
\vec r - \vec r'|^2} \nonumber \\
& & (p^i({\bf r'}) u^j + p^j({\bf r'}) u^i - (g^ {ij} - u^i u^j) 
p^c({\bf r'}) u_c) + \nonumber \\
& & {\int}_{V - \{x\}} \sqrt {g} d^3 {\bf x'} \frac {3}{|\vec r - \vec r'|^3} 
\nonumber \\
& & ({\epsilon}^{imn}s_m ({\bf r'}) u_n u^j + {\epsilon}^{jmn}s_m 
({\bf r'}) u_n u^i). \label {smat}
\end {eqnarray}
The integration is performed in the horizon of point $x$ except the point 
itself. Vector $u^i$ is in the direction of the source at $\vec r'$. The 
vectors $p^i({\bf r})$ and $s^i ({\bf r})$ are linear and angular velocity 
field densities with following definitions:
\begin {equation}
{\epsilon}^{ijk}p_{j|k} = 0, \quad \quad s^i_{|i} = 0.
\end {equation}
In general, these vectors can be obtained from matter distribution. Boltzmann 
equation relates them to non-gravitational interactions \cite{boltz}.
\\ 
Finally, $S^{ij}$ can have a purely transverse-traceless component independent 
of matter distribution and related to relic gravitational waves. Therefore 
$S^{ij} = S_{vel}^{ij} + S_{relic}^{ij}.$\\ 
Having $S^{ij}$ for a given distribution of matter and relic perturbations, we 
can now use equations 
(\ref {mtt}) to (\ref {ml}) to determine the coefficients of (\ref {phieq}).
Using approximation (\ref {delphi}) and the fact that for the 
chosen base metric $^{(3)} R = 0$, equation (\ref {phieq}) changes to:
\begin {eqnarray}
{\triangle} \delta \phi & = & - \frac {1}{8} [(M_{TT} + 2 M_{TL} + M_{L} - 
\nonumber \\
& & \frac {2}{3} K^2_{homo} + 16 \pi G T_{*homo}^*) - (7 M_{TT} + 2 M_{TL} - 
5 M_{L} + \nonumber \\
& & \frac {10}{3} K^2 + 48 \pi G T_{*homo}^*)] \delta \phi + 16 \pi \delta 
T_*^* - \frac {4}{3}K_{homo} \delta K. \label {delphieq}
\end {eqnarray}
In a homogeneous universe:
\begin {equation}
\bar g_{ij} = {\alpha}^2 \delta_{ij} \quad \mbox {and} \quad \bar K_{ij} = 
-\frac {d\alpha}{d\eta} \delta_{ij}. 
\end {equation}
From (\ref {enrcons}), one can see that:
\begin {equation}
16 \pi G \bar T_{*homo}^* - \frac {2}{3} K_{homo}^2 = 16 \pi G T_{*homo}^* - 
\frac {2}{3} K_{homo}^2 = 0.
\end {equation}
To keep $\delta \phi$ small according to our small perturbation assumption, the 
matter has to have a small average momentum. This makes all coefficients 
$M_{TT}$, $M_{TL}$ and $M_{L}$ 
of second-order and negligible. It is also another 
demonstration of power of this method. \cite{seed} also arrives at 
the same conclusion for the vorticity of a perfect fluid. Here this result 
is obtained without any assumption about type and state equation of matter. 
At first-order, the velocity perturbations contribute only to the evolution 
of the metric (equation (\ref {gevol})) and not to constraints.\\ 
With these perturbative simplifications, (\ref {delphieq}) reduces to:
\begin {equation}
{\triangle} \delta \phi = - 2 \pi G \delta \bar T_*^* + \frac {1}{6}K_{homo} 
\delta K= (\frac {5}{12} K^2 + 6 \pi G T^*_{*homo}) \delta \phi - 2 \pi G 
\delta T_*^* + \frac {1}{6}K_{homo} \delta K. \label {laplacphi}
\end {equation}
and can be solved as:
\begin {eqnarray}
\delta \phi ({\bf x},t) & = & \frac {1}{\alpha (t) ^2}\int d^3 {\bf x'} 
\frac {2 \pi G \delta \bar T_*^* ({\bf x'}, t) - \frac {1}{6}K_{homo} 
\delta K ({\bf x'}, t)}{|\vec r - \vec r'|} \nonumber \\
& = & \frac {1}{(2\pi)^3}\int d^3 {\bf k} e^{-i\alpha (t)
{\delta}_{ij}k^ix^j} \frac {2 \pi G \delta T^*_* ({\bf k}, t)- \frac {1}{6}
K_{homo} \delta K ({\bf k}, t)}{k^2 + \frac {5}{12} K_{*homo}^2 + 6 \pi G 
T^*_{*homo}}. \label {phiksol}
\end {eqnarray}
($\vec r$ and $\vec r'$ are as in (\ref {smat})). From the form of the base 
metric, $T^*_{*homo} = \bar T^*_{*homo}$.\\ 
The first expression for $\delta \phi$ shows that physical metric 
$\bar g_{ij}$ is similar to the metric for scalar perturbations alone 
(see e.g. \cite{bert}). It is the effect of special coordinate choice on 
the 3-space.\\ 
We now have all quantities necessary for definition of a set of initial data 
and solution of evolution equations (\ref {gevol}) and (\ref {kevol}). 
The essential characteristic of linearized constraints is that equations 
for $W^i$ and $\phi$ are decoupled. In fact, for all order of 
perturbations, $\delta \phi$ decouples from (\ref {weq1}) and therefore, this 
method is easily applicable to higher-order perturbative calculations.

\section {Non-Flat Cosmogonies}
For the general case of a space-time with constant curvature, we can use a 
flat base metric as before. The perturbative expansion of $\phi$ will 
take the following form:
\begin {eqnarray}
\phi ({\bf x}, t) & = & {\phi}_0 (1 + \delta \phi ({\bf x}, t)).
\label {dphir}\\
{\phi}_0 & = & \frac {1}{(1 + \frac {\hat K}{4} r^2)^\frac {1}{2}}. \\
r^2 & = & {\delta}_{ij} x^i x^j \\
\bar g_{ij} & = & {\alpha}^2 (t) ({\gamma}_{ij} + h_{ij}) \nonumber \\ 
& = & {\alpha}^2 (t) {\phi}^4 ({\bf x}, t) {\delta}_{ij} = {\alpha}^2 (t) 
{\gamma}_{ij} (1 + 4 \delta \phi ({\bf x}, t))\\
{\gamma}_{ij} & = & \frac {{\delta}_{ij}}{(1 + \frac {\hat K}{4} r^2)^\frac 
{1}{2}}.\label {gdef1}
\end {eqnarray}
$\hat K$ is the constant curvature of the homogeneous universe. After 
performing a variable change from $x$ to $x'$ as before, the equation for 
$W^i$ will become:
\begin {equation}
{\partial}'_j [\frac {{\partial}'^i W^j + {\partial}'^j W^i - \frac {2}{3}
{\delta}^{ij} {\partial}'_c W^c}{(1 + \frac {\hat K'}{4} r'^2)^{3}}] = 
- 8 \pi G T^{*i} + \frac {{\partial}'^i K}{3}. \label {weqr}
\end {equation}
where $\hat K' = \frac {\hat K}{\alpha (t)^2}$. To solve this equation, 
we assume, without loss of generality, that $W^i$ can be decomposed to:
\begin {eqnarray}
W^i & = & \psi V^i - {\partial}'^i U. \label {vu} \\
\psi & = & (1 + \frac {\hat K'}{4} r'^2)^{3}.
\end {eqnarray}
Putting (\ref {vu}) into (\ref {weqr}) gives a system of equations for $V^i$ and $U$:
\begin {eqnarray}
{\partial}'^i{\partial}'_j V^j + {\partial}'^j{\partial}'_j V^i - 
\frac {2}{3} {\delta}^{ij} {\partial}'_j{\partial}'_c V^c & = & 
- 8 \pi G T^{*i}({\bf x'}, t) + \frac {K^{|i}}{3}. \label {veq}\\
{\partial}'^4 U & = &\frac {3}{4}{\partial}'_i{\partial}'_j F^{ij}. \label {ueq}
\end {eqnarray}
\begin {equation}
F^{ij} \equiv V^j {\partial}'^i \psi + V^i {\partial}'^j \psi + 
\frac {2}{3} {\delta}^{ij} V^c {\partial}'_c \psi. \label {fdef}
\end {equation}
Equation (\ref {veq}) is exactly the same as equation (\ref {weq3}) and 
therefore its solution is the same as (\ref {wsol}). In (\ref {ueq}), the 
right hand side is known once (\ref {veq}) is solved and the solution of 
(\ref {ueq}) is:
\begin {equation}
U ({\bf x}, t) = -\frac {1}{(2\pi)^3}\int d^3k e^{-i\alpha (t){\delta}_{ij}
k^ix^j} \frac {3k_ik_j F^{ij}(k)}{4 k^4}.
\end {equation}
Equation (\ref {delphieq}) changes to:
\begin {equation}
{\triangle} \delta \phi = (\frac {5}{12} K_{homo}^2 + 
6 \pi G T^*_{*homo}) {\phi}_0^4 \delta \phi - 2 \pi G {\phi}_0^{-3} 
\delta T_*^* - \frac {4}{3} {\phi}_0^5 K_{homo} \delta K. \label {phiphieq}
\end {equation}
To solve this equation, one can expand $\delta \phi$ and $\delta T^*_*$ with 
respect to Spherical Harmonic functions:
\begin {eqnarray}
\delta \phi & = & \sum_{ml} a_{ml}(r) Y_{ml} (\theta, \varphi) \quad \quad 
\delta T^*_* = \sum_{ml} T_{ml}(r) Y_{ml} (\theta, \varphi) \nonumber \\
\delta K & = & \sum_{ml} K_{ml}(r) Y_{ml} (\theta, \varphi)
\end {eqnarray}
This results to the following equation for $a_{ml}$:
\begin {eqnarray}
r^2 \frac {d^2 a_{ml}}{dr^2} + 2r \frac {da_{ml}}{dr} + (l(l+1) - r^2 A (r)) a_{ml} 
& = & \nonumber \\
\frac {d}{dr}(r^2 \frac {da_{ml}}{dr}) +  (l (l + 1) - r^2 A (r)) a_{ml} 
& = & r^2 (B(r) T_{ml} + D(r) K_{ml}). \label {req}
\end {eqnarray}
\begin {eqnarray}
A (r) & \equiv & {\alpha}^2 (t)(\frac {5}{12} K_{homo}^2 + 
6 \pi G T^*_{*homo}){\phi}_0^4 \\ 
B (r) & \equiv & - 2 \pi G {\alpha}^2 (t) {\phi}_0^{-3} \\
D (r) & \equiv & - \frac {4}{3} {\phi}_0^5 K_{homo}.
\end {eqnarray}
This equation has a polynomial homogeneous solution. For 
simplifying our notation, in the following we ignore $m$ and $l$ indexes. We 
assume:
\begin {equation}
a (r) = \sum_n d_n r^n.
\end {equation}
for the homogeneous solution. Replacing $a_{ml}$ in (\ref {req}) with this 
expansion, one obtains the following 
recurrent expression for the coefficients $d_n$:
\begin {equation}
d_n = \sum_{i=0}^8 \frac {C_{2i}}{C_1}[(n-2i+2)(n-2i + 3) + l (l +1)]d_{n-2i+2}. 
\label {dn}
\end {equation}
\begin {eqnarray}
C_1 & = & {\alpha}^2 (t)(\frac {5}{12} K_{homo}^2 + 6 \pi G T^*_{*homo}). \\
C_{2i} & = & \frac {8!(\frac {\hat K}{4})^i}{i! (8 - i)!}.
\end {eqnarray}
One can construct two independent solutions for two boundary conditions:
\begin {eqnarray}
& 1) & |a| < \infty \quad \mbox {for} \quad r \rightarrow 0 \quad 
\Longrightarrow \quad d_n = 0, \quad n < 0. \\
& 2) & |a| < \infty \quad \mbox {for} \quad r \rightarrow \infty \quad 
\Longrightarrow \quad d_n = 0, \quad n > -16.
\end {eqnarray}
As the derivative terms of (\ref {req}) are complete, a Green function can 
be found for this equation (\cite {eqhand}). We call the above homogeneous 
solutions $a^{(1)}(r)$ and $a^{(2)}(r)$. The Green function and the complete 
solution of $a_{ml}$ are determined as:
\begin {eqnarray}
G (r;r') & = & \left \{ \begin {array}{ll} \frac {a^{(1)}(r)a^{(2)}(r')}{r'^2 \omega (r')} & 0 \leq r \leq r', \\
\frac {a^{(1)}(r')a^{(2)}(r)}{r'^2 \omega (r')} & r' \leq r < \infty 
\end {array} \right. \\
\omega (r) & = & a^{(1)}(r)\frac {da^{(2)}}{dr} - a^{(2)}(r)\frac 
{da^{(1)}}{dr}. \\
a_{ml} (r) & = & \int dr' r'^2 (B (r') T_{ml} + D (r') K_{ml}) G (r;r') .
\end {eqnarray}
This completes the solution of constraints for perturbed constant curvature 
spaces.

\section {Matching Condition}
The results obtained in previous sections can be applied to any 
perturbative cosmological context. An straightforward and interesting 
application is the determination of initial metric and extrinsic curvature 
perturbations in a universe filled with species which have a distribution 
$f({\bf x}, t, {\bf p})$ in their classical phase space. If they have mutual 
interactions, one usually has to solve 
the Einstein-Boltzmann equations numerically. Division of these equations to 
scalar, vector and tensor components increases significantly the amount of 
equations to be solved and consequently the calculation time. This issue and 
the detail of calculation will be discussed elsewhere \cite {pots}, 
\cite {uhdm}.\\
Here we use constraint solutions to find a general expression for matching 
condition on a surface of phase transition. This issue has been already 
discussed in \cite {definit}, \cite {definit1}, \cite 
{matchcons}. We show that on the light of results obtained 
above, the matching becomes trivial and its physical interpretation 
more transparent. For simplicity, in this section we consider only the case 
of a flat cosmology. Using formalism presented in the previous section, 
the extension to a curved space-time is straightforward.\\
The matter in the early universe has gone through a few number of phase 
transitions. One of the consequences of these transitions is the formation of 
topological defects which could have been the source of initial perturbations 
(if not completely, at least partially \cite {infseed}). To study the 
evolution of fluctuations after their formation, it is usually assumed that 
transition was very fast i.e. its duration was much shorter than evolution 
time until matter-radiation equilibrium. In this case, a phase transition 
approximately defines an equal time 3-space in the space-time.\\
The perturbation of the energy-momentum tensor after transition is written as:
\begin {equation}
\delta \bar T_{\mu \nu} = \delta \bar T_{\mu \nu}^{\mbox {rad}} + 
{\Theta}_{\mu \nu} \label {thetadecom}
\end {equation}
where $\delta \bar T_{\mu \nu}^{\mbox {rad}}$ is the perturbation in the 
ordinary matter (assumed to be relativistic), and ${\Theta}_{\mu \nu}$ is the 
energy-momentum tensor of defects. In stiff approximation, it is assumed that 
${\Theta}_{\mu \nu}$ evolves separately and is treated as an external source.\\
The power spectrum of defects after their formation can only be determined by numerical 
simulations \cite {defdens}. It needs a complete simulation of quantum processes 
during phase transition and their decoherence which leads to macroscopic 
classical behavior of defects. Such a simulation is not yet available and consequently, 
it is necessary to use 
phenomenological arguments to fix the initial conditions in simulations of 
perturbation evolution in presence of defects \cite {seed}, \cite {pen}, \cite 
{defsim}.\\
It is usually assumed that the effect of defect formation on matter and 
radiation is tiny and consequently perturbative. For studying the evolution of 
large scale (wavelength) perturbations which are important for the formation 
of large structures today, the lowest terms in the expansion of the power 
spectrum of defects is enough and give an analytical expression for initial 
perturbations \cite {definit1}. 
Some of amplitudes of the expansion terms can be fixed by using the conservation of 
${\Theta}_{\mu \nu}$, i.e ${\Theta}^{\mu \nu}_{;\nu} = 0$. Others are usually 
fixed by some physical arguments, like causality, and/or a matching condition 
that relates a physical quantity in two parts of the space-time separated 
by the phase transition surface. Matching along with other choices define a 
model for initial perturbations. These conditions must be at least consistent 
and treated together.\\
For physical reasons like conservation of energy-momentum tensor, the phase 
transition surface is assumed to be a surface of constant density. However, 
this definition is ambiguous. In a flat space-time before 
phase transition, a constant density 3-space is flat. After transition, in 
general, the constant density surface is not equal-time and flat (keeping 
the same coordinate definition). To remove this ambiguity, we assume that 
the isomorphism between homogeneous and perturbed space-time manifolds 
${\mathcal M}_{homo}$ and ${\mathcal M}$, maps $\Sigma$, the transition 
surface, to a flat equal-time 
(thus constant density) 3-space. This means that at ${\mathcal O} (0)$ the 
density is constant but not at ${\mathcal O} (1)$ (It is the general case. 
Evidently, it is always possible to redefine time parameter such that 
$\Sigma$ becomes an equal time and constant density surface).\\
By definition, the space time must be differentiable everywhere including on 
the initial 3-space. This means that $a (t)$, $\bar g_{ij}$ and $\bar K_{ij}$ 
must be continuous \cite {definit}:
\begin {equation}
[\bar g_{ij}]_{\pm} = 0 \quad \mbox {and} \quad [\bar K_{ij}]_{\pm} = 0. 
\label {match}
\end {equation}
where for any quantity $Q$, $[Q]_{\pm} \equiv \lim_{\epsilon \rightarrow 0} 
Q (t_{PT} + \epsilon) - Q (t_{PT} - \epsilon)$. Variable $t_{PT}$ is the phase 
transition comoving time. Note that if these conditions are satisfied in one 
gauge, a continuous gauge transformation fulfills them too. Therefore in this 
sens, they are gauge invariant.\\
The relation (\ref {match}) is used as the 
matching condition on the phase transition surface. Here we apply the solution 
of constraints (\ref {enrcons}) and (\ref {momcons}) to the matching conditions, 
and determine the unconstrained quantities $S^{ij}$ and $K$ on the initial 
3-space.\\
From (\ref {lw}) and (\ref {wsol}):
\begin {eqnarray}
(LW)^{ij}({\bf k}, t) & = & \frac {2 \pi G}{k^4}[4k^2 (k^i T^{*j} + 
k^j T^{*i}) - 2 k_c T^{*c} (k^i k^j + k^2 g^{ij}) - \nonumber \\
 & & \frac {k^2 \delta K}{4 \pi G} (k^i k^j - \frac {k^2 g^{ij}}{3})]. 
\label {lwsol}
\end {eqnarray}
and from (\ref {sijtrans}), (\ref {lwtrans}), and (\ref {kbdecom}):
\begin {eqnarray}
\bar K^{ij} & = & (S^{ij}_{homo} + (LW)^{ij}_{homo} + K_{homo} g^{ij}) + 
\delta S^{ij} + \delta (LW)^{ij} - \nonumber \\
 & & (10 S^{ij}_{homo} + 4 (LW)^{ij}_{homo} + 
\frac {4}{3} K_{homo} g^{ij}) \delta \phi + g^{ij} \delta K. \label {kbar}
\end {eqnarray}
Equation (\ref {laplacphi}) gives $\delta \phi$ as a function of  
$\delta \bar T^*_*$:
\begin {equation}
\delta \phi = \frac {2 \pi G \delta \bar T^*_* + \frac {H}{2} \delta K 
({\bf k}, t)}{k^2} \quad \quad H \equiv \frac {\dot \alpha}{\alpha}. 
\label {phival}
\end {equation}
Dots denote derivative with respect to comoving time $t$. Before phase 
transition, Universe is homogeneous and the physical metric and extrinsic 
curvature are 
equal to their bare counterparts. The matching condition for induced metric 
(\ref {match}) and definition of physical and bare metric (\ref {gdef}) and 
(\ref {baremet}) leads to $\delta \phi = 0$ and thereafter $\bar g_{ij} = g_{ij} = {\alpha}^2 (t) {\delta}_{ij}$ and:
\begin {equation}
\delta K = - \frac {4 \pi G}{H} \delta \bar T^*_*. \label {ksol}
\end {equation}
The matching condition for extrinsic curvature $\bar K^{ij}$ is 
$\delta K^{ij} = 0$. In consequence, its three independent components i.e. 
$\delta K$, $(LW)^{ij}$ (it is ${\mathcal O} (1)$), and $S^{ij}$ must 
separately be zero. From (\ref {ksol}) and (\ref {lwsol}):
\begin {equation}
\delta \bar T_*^* = T^{*i} = 0. \label {t00sol}
\end {equation}
(\ref {t00sol}) means that there is no total density or velocity perturbation on 
the initial 3-space. If matter is a perfect fluid or a scalar field, i.e. 
without viscosity, $\delta \bar T _{ij} = 0$.\\
The above results show that the matching condition (\ref {match}) is sufficient 
for determination of the initial value of metric and extrinsic curvature and 
leads to an isocurvature initial condition for large wavelengths 
perturbations for matter without viscosity (e.g. a mixture of perfect fluid and a scalar field). No degree of 
freedom rests to be chosen or fixed by other physical arguments.\\
Some authors (e.g. \cite {pen} and \cite {imitinfl}) use an ordinarily 
(in contrast to covariantly) conserved pseudo-energy-momentum tensor to define 
a matching condition. In terms of induced metric and extrinsic curvature of 
the initial 3-space, this tensor has the following form:\\
\begin {equation}
{\tau}_{\mu \nu} = \left [
\begin {array}{cc}
\delta \bar T_{00} + \frac {H}{4 \pi G} \delta \bar K  &  
\delta \bar T_{0i} \\ 
\delta \bar T_{0i}  &  \delta \bar T_{ij} - \frac {{\alpha}^2 H}
{4 \pi G} \left (H h_{ij} - \delta K_{ij} + {\delta}_{ij} \delta K \right )\\
\end {array}
\right ] \label {taumunu}
\end {equation}

If we impose the continuity of $h_{ij}$ and $\bar K_{ij}$ as before, from 
(\ref {ksol}) and (\ref {t00sol}) one can trivially conclude that 
$\tau_{\mu \nu}$ conservation condition is fulfilled. In a flat space-time 
the ${\bf k} \neq 0$ modes of $\tau_{\mu \nu}$ are zero. 
\cite {pen} and \cite {imitinfl} use this property as matching condition 
and set $\tau_{00} = 0$ and $\tau_{0i} = 0$. In contrast to the first 
prescription, these conditions don't fix all the independent degrees of 
freedom. For fixing the rest, Pen et al. choose a relation between density 
perturbation of radiation and dark matter (based on the assumption of a white 
noise power spectrum at superhorizon scales). \cite {imitinfl} chooses these 
quantities such that at superhorizon scale there is not any perturbation in 
the ratio of different components of the plasma. It is equivalent to an 
unperturbed initial condition.\\
In \cite {matchcons}, first (\ref {match}) is used as matching condition and a 
series of relations between components of metric, extrinsic curvature and 
energy-momentum tensor of defects is obtained. Then it is shown that to first 
order, these conditions is equivalent to $\tau_{00} = 0$ i.e. the condition 
used by \cite {pen} and \cite {imitinfl} for matching. From (\ref {ksol}), 
(\ref {t00sol}) and (\ref {taumunu}) one can immediately and without lengthy 
demonstration of \cite {matchcons} conclude that continuity of $\bar g_{ij}$ 
and $\bar K_{ij}$ imposes $\tau_{00} = 0$. But the inverse is not true, i.e. 
if $\tau_{00} = 0$ and $\tau_{0i} = 0$, it does not necessarily mean that 
there is no total initial perturbation even for a matter without viscosity. 
Therefore, equivalence of two matching 
prescriptions is one directional and model dependent. Specially, for some 
models like perfect fluid and scalar field, the purely geometric matching of 
Deruelle et al. completely fixes the initial value of the spectrum 
((\ref {ksol}) and (\ref {t00sol})), but not the other prescription. 
The advantage of the method 
presented here is that it is compact and independent of the details of the 
model. This makes the interpretation of the results easier and more transparent.
\\
Finally, we rise the following question: How physically meaningful is matching? A matching prescription is used in the circumstances that not enough physical 
information about the model is available to fix the initial 
value of geometry and spectrum. However, even a purely geometrical and 
physically well motivated prescription of Deruelle et al. leads automatically 
to an isocurvature perturbations in a mixture of perfect fluid and 
scalar field. This can be a too much simplification of the reality. One 
interpretation of this result is that any physical process needs a finite time 
to happen and a complicate process like defect formation can not be replaced by 
a geometric matching on a 3-space. In this situation, it is probably more 
reasonable to use some phenomenological arguments for choosing the initial 
conditions.\\
In conclusion, York method for separation of dependent and independent degrees 
of freedom in Einstein equations is used to solve analytically the constraint 
equations for small perturbations in space-times with constant curvature. The 
solution is independent of details of the matter model. The results is applied 
to two commonly used matching prescriptions and it is shown that their 
equivalence is partial and model dependent.

\end{document}